\documentclass[10pt]{article}
\usepackage{graphicx}
\usepackage{amsfonts}
\begin{document}

\def\noi{\noindent}
\def\np{\not{P}}
\def\udt#1{$\underline{\smash{\hbox{#1}}}$}
\def\F{\Cal{F}}
\def\D{\Cal{D}}
\def\X{\Cal{X}}
\def\wt{\widetilde}
\def\ov{\overline}
\centerline{\udt{\bf ON THE COVARIANT QUANTIZATION OF GREEN-SCHWARZ}}
\centerline{\udt{\bf SUPERSTRING AND BRINK-SCHWARZ SUPERPARTICLE}}
\vskip .2cm
\centerline{M. Caicedo, M Lled\'o {\udt{A. Restuccia}} and J. Stephany}
\vskip .2cm
\centerline{Dept. of Physics, Simon Bolivar University, Caracas}
\vskip .4cm

The covariant quantization of the interacting Green-Schwarz \cite{[1]}
Superstring (GSSS) is one of the interesting problems to solve in String
Theory.The formulation of Superstrings a la Green-Schwarz manifestly
supersymmetric on the target manifold has several advantages over the NRS \cite{[2]}
formulation. In fact, the latest is only a perturbative approach to the
interacting Superstring Theory. Moreover, supersymmetry on the target  manifold
is obtained only after summation over spin structures. However this  problem
for odd spin structures has not been completely understood.

The Green-Schwarz second quantized action for superstrings has only been
consistently formulated in the Light Cone Gauge \cite{[3]}. In spite of the fact
that several attempts have been performed, the covariant formulation of the
problem has not been obtained. The difficulty to start with, has been that the
first class constraints associated to gauge symmetries of the first
quantized theory appear mixed with second class ones and no local, Lorentz
covariant, approach to extend them to a set of only first class constraints
has been yet developed.The same problem appears in the covariant quantization
of the  Brink-Schwarz \cite{[4]} Superparticle (BSSP) which describes the zero mode
structure of the GSSP.

In order to circumvent the problem presented by the mixing of first and second
class  constraints in a covariant treatment other actions for the description
of the  Superparticle, following original ideas of Siegel were proposed. The
so-called  Siegel Superparticle, SSP \cite{[5]}, does not have the same number of
degrees of  freedom as the BSSP. It corresponds to ignore the second class
constraints of  the BSSP formulation, leaving only the first class constraints
which may be  covariantly projected from the original BSSP set of constraints.
The fact that SSP  has a number of degrees of freedom different from BSSP is a
consequence of the non trivial restrictions imposed by the second class
constraints.

The Modified Siegel Superparticles, MSSPI \cite{[6]} and MSSPII \cite{[7]} have
the same physical spectrum as the BSSP, allowing a formulation in
terms of first class constraints only. The formulations however
are given in terms of {\udt{irregular}} constraints, in
distinction from the original BSSP formulation. If the constraints
of a dynamical system are expressed as the kernel of a map  $\phi$
between Banach manifolds,
\begin{displaymath}
\phi :M\to N
\end{displaymath}
and if for some $m$ belonging to the kernel of $\phi$ the induced tangent map
at  $m$, $T_m\phi$, is not surjective we call the constraint irregular.

If $\phi$ is a regular constraint $\phi^{-1}(0)$ always define a
submanifold. If $\phi$ is irregular, $\phi^{-1}(0)$ is not necessarily a
submanifold. For irregular constraints the Lagrange multiplier theorem does
not in general apply. In fact one of the hypothesis is the regularity of the
constraints. This implies that the generalized canonical quantization which
introduces all constraints into the action using Lagrange multipliers is not
in general valid. In fact the effective action of a constrained theory with
first class irregular constraints is well defined only in a gauge implemented
by a gauge condition on the associated Lagrange multipliers. In particular
this restriction allows a covariant gauge condition. However, the action on a
canonical gauge is ill defined. Consequently the off-shell reduction to the
physical modes, which is one of the main arguments to prove unitary of the
S-matrix by  functionals methods, cannot be proven.

A general discussion on the quantization for irregular systems is given in \cite{[8]}.
The presence of irregular constraints in all the Modified Siegel Superparticle
and Superstring actions does not allow, because of the previous arguments, the
direct application of the Batalin-Vilkovisky approach to quantize the problem,
and so far have failed to solved the problem of the covariant quantization.

For the same reasons the off-shell nilpotent BRST charge for the
Superparticle \cite{[9]} has been obtained only in a indirect way which does not give
any insight into the geometric structure of the theory nor any idea on the
generalization to the Superstring problem.

In this work we present a direct alternative approach to  the
covariant quantization of the BSSP using infinite  auxiliary
fields. At any truncated level the theory is not Lorentz
covariant.  The covariance is obtained only after the introduction
of the infinite  auxiliary fields. It leads directly to BRST
charge previously found in \cite{[9]} and  to the Kallosh action \cite{[10]}.
The formulation is based upon a general canonical  approach for
dynamical systems restricted by reducible first and second  class
constraints \cite{[11],[12]}. In this approach the phase space is
extended to a larger manifold where all extended constraints are
first class. By an appropriate gauge fixing one may reduce the
functional integral to a functional integral on the original
constrained manifold, with the correct functional measure. It is
an off-shell approach allowing the systematic construction of the
off-shell nilpotent BRST charge and of the BRST invariant
effective action.

The first order action for the ten dimensional BS superparticle is
\begin{equation}
\label{1}
S=<P_\mu \partial_\tau \chi^\mu +\ov{\xi} \np
\partial_\tau \xi+eP^2>
\end{equation}
where $e$ is a Lagrange multiplier associated to the constraint
\begin{equation}
\label{2}
P^2=0\ \ \ \ \ .
\end{equation}

Let $\eta$ be the momenta canonically conjugate to $\xi$. Since the action
(\ref{1}) is first order in $\partial_\tau\xi$ its dynamics is restricted by,
\begin{equation}
\label{3}
\phi =\eta -\np \xi =0 \ \ \ \ \ .
\end{equation}

The canonical Hamiltonian action of the system is
\begin{equation}
\label{4}
S=<P_\mu \partial_\tau \chi^\mu +\ov{\eta} \partial_\tau \xi
+eP^2+ \ov{\psi}(\eta -\np \xi )>
\end{equation}
where
$\ov{\psi}$ are Lagrange multipliers associated to the constraints
(\ref{3}).

Constraints (\ref{3}) are a combination of first and second class ones.
We enlarge the phase space by introducing the following auxiliary fields
\begin{displaymath}
\Phi_1 =\eta_1+\np \xi_1\\
\ov{\Phi}_1 =\eta_1-\np \xi_1 \ \ \ \ \ .
\end{displaymath}

The enlarged constraints are in this case
\begin{equation}
\label{5}
\wt{\phi}_0=\eta -\np \xi +\Phi_1\ \ \ \ \ .
\end{equation}

It is necessary to add the restriction
\begin{equation}
\label{6}
\ov{\Phi}_1^\top =0\ \ \ \ \ .
\end{equation}
where $\top$ is the transverse projection in the sense \cite{[11]}.

The factor $det\{\ov{\Phi}^\top_1,\ov{\Phi}^\top_1\}^{1/2}$ in the
measure of the functional integral is in principle as problematic
as the factor $det\{\phi^\top ,\phi^\top \}^{1/2}$ in the  direct
approach. Nevertheless we observe that constraint (\ref{6}) is
equivalent to a reducible constraint
\begin{equation}
\label{7}
\ov{\Phi}_1 = 0,\\
\np \ov{\Phi}_1|_{{first class}} =\np \eta_1 \equiv 0 .
\end{equation}

We iterate now the process and introduce $\xi_2$, $\eta_2$ and $\omega_2$.
We obtain again
\begin{equation}
\label{8}
\omega_2 =-2\np \\
\Phi_2 =\eta_2+\np \xi_2 \\
\ov{\Phi}_2 =\eta_2-\np \xi_2
\end{equation}
and we have the new constraints
\begin{equation}
\label{9}
\wt{\phi}_1 =\ov{\Phi}_1+ \Phi_2 \\
\ov{\Phi}^\top_2 =0
\end{equation}

For the same reason as above we take instead of (\ref{9}) the reducible constraint
\begin{equation}
\label{11}
\ov{\Phi}_2 =0\\
\np \ov{\Phi}_2|_{{first class}} =\np \eta_2 \equiv 0
\end{equation}
and continue the process. After $\ell$ steps we have
\begin{equation}
\label{12}
\wt{\phi}_{i-1} =\ov{\Phi}_{i-1}+\ov{\Phi}_i\ \ \ \ i=1,\cdots ,\ell\\
\ov{\Phi}^\top_{\ell} =0
\end{equation}
with $\ov{\Phi}_0 \equiv \phi$.

At this level the classical action may be written in terms of the canonical
variables in the form
\begin{equation}
\label{13}
S_\ell =<P_\mu \dot{x}^\mu +\sum_{i=0}^\ell \eta_i \dot{\xi}^i+
\lambda P^2+\sum_{i=0}^{\ell -1}\ov{\psi}^i\np \eta_i +
\sum^{\ell}_{i=0}\lambda^i\wt{\phi}_{i-1}>.
\end{equation}
subject to the second class constraints
\begin{displaymath}
\ov{\Phi}^\top_\ell =0.
\end{displaymath}

Here $\wt{\phi}_\ell$ in (\ref{13}) is irreducible,the other constraints being
infinite reducible. At each level $\ell$ the formulation (\ref{13}) is not Lorentz
covariant due to the  transverse projection the second equation.

In order to avoid this problem we may introduce infinite auxiliary fields. We
then obtain
\begin{equation}
\label{14}
S_\infty =<P_\mu \dot{x}^\mu +\sum_{i=0}^\infty \eta_i \dot{\xi}^i
+ \lambda P^2+\sum_{i=0}^{\infty}\ov{\psi}^i\np \eta_i +
\sum^{\infty}_{i=0}\lambda^i\wt{\phi}_{i-1}>.
\end{equation}
which was first obtained by Kallosh in \cite{[10]}.

The effective action associated to (\ref{14}) may be truncated at any
level $\ell$ by imposing appropriate gauge fixing conditions and
the effective action associated to (\ref{13}) is regained. In the limit
case $\np$ $\eta_i$ and $\wt{\phi}_i$ $i=0,\cdots ,\infty$ are
regular infinite reducible first class constraints. This  allows
the systematic Batalin-Fradkin construction of the off-shell
nilpotent BRST charge. The result as was shown by Kallosh in ref.
\cite{[10]} match with the BRST operator with the correct cohomology for
the BSSP.

The superparticle action on both an $N=1$ and $N=2$ superworldline was
discovered by Sorokin, Tkach and Volkov \cite{[13]}. We shall refer them as STVSP1 and
STVSP2 respectively.

The STVSP1 contains as one of the field equation the same constraint (3) as in
the BSSP. The correct quantization of it follows in the same lines as we have
described the BSSP.

The STVSP2 contains an extension of (\ref{3}) to a more general phase
space which coincides with the extension proposed in \cite{[14]}. The set
of constraints proposed in \cite{[14]} are first class only and have two
stages of reducibility. However the twistorial approach of \cite{[14]}
doesn't get rid of the second class constraints present in the
Superstring case.

An extension of the approach presented here, for the covariant treatment of
the Superparticle, has been performed which in particular generalizes the
twistorial approach of \cite{[14]}. Its application to the GSSS is under study. We
hope to report on it soon.

\end{document}